\documentclass[onecolumn]{aastex61}

\begin{document}

\title{Detecting Proxima b's atmosphere with JWST targeting CO$_2$ at 15 micron using a high-pass spectral filtering technique}

\correspondingauthor{Ignas Snellen}
\email{snellen@strw.leidenuniv.nl}

\author{I.A.G. Snellen}
\affil{Leiden Observatory, Leiden University, Postbus 9513, 2300 RA Leiden, The Netherlands}
\author{J.-M. D\'esert}
\affil{Anton Pannekoek Institute for Astronomy, University of Amsterdam, P.O. Box 94249, 1090 GE Amsterdam, the Netherlands}
\author{L.B.F.M. Waters}
\affil{SRON Netherlands Institute for Space Research, Sorbonnelaan 2, 3584 CA Utrecht, the Netherlands}
\affil{Anton Pannekoek Institute for Astronomy, University of Amsterdam, P.O. Box 94249, 1090 GE Amsterdam, the Netherlands}
\author{T. Robinson}
\affil{Department of Astronomy and Astrophysics, University of California, Santa Cruz, CA 95064, USA}
\affil{NASA Sagan Fellow}
\author{V. Meadows}
\affil{Astronomy Department, University of Washington, USA}
\author{E.F. van Dishoeck}
\affil{Leiden Observatory, Leiden University, Postbus 9513, 2300 RA Leiden, The Netherlands}
\author{B.R. Brandl}
\affil{Leiden Observatory, Leiden University, Postbus 9513, 2300 RA Leiden, The Netherlands}
\affil{Delft University of Technology, Faculty of Aerospace Engineering, Kluyverweg 1, 2629 HS Delft, The Netherlands}
\author{T. Henning}
\affil{Max-Planck-Institute for Astronomy, Koenigstuhl 17, 69117 Heidelberg, Germany}
\author{J. Bouwman}
\affil{Max-Planck-Institute for Astronomy, Koenigstuhl 17, 69117 Heidelberg, Germany}
\author{F. Lahuis}
\affil{SRON Netherlands Institute for Space Research, Sorbonnelaan 2, 3584 CA Utrecht, the Netherlands}
\author{M. Min}
\affil{SRON Netherlands Institute for Space Research, Sorbonnelaan 2, 3584 CA Utrecht, the Netherlands}
\author{C. Lovis}
\affil{Observatoire de Gen\`eve, Universit\'e de Gen\`eve, 51 chemin des Maillettes, 1290 Versoix, Switzerland}
\author{C. Dominik}
\affil{Anton Pannekoek Institute for Astronomy, University of Amsterdam, P.O. Box 94249, 1090 GE Amsterdam, the Netherlands}
\author{V. Van Eylen}
\affil{Leiden Observatory, Leiden University, Postbus 9513, 2300 RA Leiden, The Netherlands}
\author{D. Sing}
\affil{School of Physics, University of Exeter, Exeter, UK}

\author{G. Anglada-Escud\'e}
\affil{School of Physics and Astronomy, Queen Mary University of London, 327 Mile End Road, London E1 4NS, UK} 
\author{J.L. Birkby}
\affil{Harvard-Smithsonian Center for Astrophysics, 60 Garden Street, Cambridge MA 02138, USA}
\affil{Anton Pannekoek Institute for Astronomy, University of Amsterdam, P.O. Box 94249, 1090 GE Amsterdam, the Netherlands}\author{M. Brogi}
\affil{Center for Astrophysics and Space Astronomy, University of Colorado at Boulder, Boulder, CO 80309, USA}
\affil{NASA Hubble Fellow}

\begin{abstract}
Exoplanet Proxima b will be an important laboratory for the search for extraterrestrial life for the decades ahead. Here we discuss the prospects of detecting carbon dioxide at 15 $\mu$m using a spectral filtering technique with the Medium Resolution Spectrograph (MRS) mode of the Mid-Infrared Instrument (MIRI) on the James Webb Space Telescope (JWST). At superior conjunction, the planet is expected to show a contrast of up to 100 ppm with respect to the star. At a spectral resolving power of R=1790--2640, about 100 spectral CO$_2$ features are visible within the 13.2-15.8 $\mu$m (3B) band, which can be combined to boost the planet atmospheric signal by a factor 3--4, depending on the atmospheric temperature structure and CO$_2$ abundance. If atmospheric conditions are favorable (assuming an Earth-like atmosphere),  with this new application to the cross-correlation technique carbon dioxide can be detected within a few days of JWST observations. However, this can only be achieved if both the instrumental spectral response and the stellar spectrum can be determined to a relative precision of $\leq$1$\times 10^{-4}$ between adjacent spectral channels.
Absolute flux calibration is not required, and the method is insensitive to the strong broadband variability of the host star. Precise calibration of the spectral features of the host star may only be attainable by obtaining deep observations of the system during inferior conjunction that serve as a reference. The high-pass filter spectroscopic technique with the MIRI MRS can be tested on warm Jupiters, Neptunes, and super-Earths with significantly higher planet/star contrast ratios than the Proxima system. 
\end{abstract}

\keywords{}

\section{Introduction} 

The discovery of the exoplanet Proxima b through long-term radial velocity monitoring \citep{Anglada2016} is exciting for two reasons. First, it confirms that low-mass planets are very common around red dwarf stars, a picture that was already emerging from both transit and radial velocity surveys \citep{Berta2013,Dressing2015}. Second, the proximity of this likely temperate rocky planet at a mere 1.4 parsec from Earth makes it most favourable for atmospheric characterisation, making Proxima b an important laboratory for the search for extraterrestrial life for the decades ahead.  

Proxima b is found to orbit its host star in 11.2 days, placing it at an orbital distance of 0.0485 au.  Since the luminosity of the host star is only 0.17\% of that of our Sun, the level of stellar energy the planet receives is 30\% less than the Earth, but nearly 70\% more than Mars. This means that in principle it could have surface conditions that sustain liquid water - generally thought as a prerequisite for the emergence and evolution of biological activity \citep[e.g.][]{Kasting1993, Kopparapu2013}. Although other habitats can be envisaged outside the so called 'habitable zone' , such as under the icy surface of Jupiter's moon Europa \citep[e.g][]{Reynolds1983, Kargel2000}, it is rather unlikely that signs of biological activity under such conditions could be detected in extrasolar planet systems \citep{Lovelock1965, Segura2005}.

It is highly debatable whether Earth-mass planets in the habitable zones of red dwarf stars, such as Proxima b, could sustain or have ever sustained life. First, it is expected that the pre-main sequence of red dwarf stars lasts up to a billion years during which the stellar luminosity is significantly higher than during the main-sequence lifetime of the star. This means that the planet will have had a significantly hotter climate early on, during which it may have lost most or maybe all of its potential water content \citep{Ramirez2014, Luger2015}. Second, Proxima, as a large fraction of red dwarf stars, is a flare star that actively bombards the planet atmosphere with highly energetic photons and particles \citep{Khodachenko2007, Lammer2007}, possibly causing a large fraction of the planet atmosphere to be lost. Thirdly, planets in the habitable zone of red dwarf stars are expected to be tidally locked, and may be synchronously rotating -- always faced with the same side to the host star \citep[e.g.][]{Ribas2016,Barnes2016}. It is not clear whether a habitable climate can be sustained with such a configuration \citep[e.g.][]{Kite2011}. However, several theoretical endeavours, also in the wake of the Proxima b discovery, show that despite these possible drawbacks the planet may well host an atmosphere with liquid water on its surface \citep{Tarter2007, Ribas2016, Turbet2016}. 

Several studies have investigated the potential detectability of Earth-like atmospheres of planets orbiting late M-dwarfs using high-dispersion spectroscopy. \citet{Snellen2013} calculated whether the transmission signature of molecular oxygen of a twin earth-planet in front of a mid-M dwarf could be observed using high-dispersion spectroscopy \citep[see also][]{Rodler2014}  and showed that it would require a few dozen transits with the European Extremely Large Telescope (E-ELT) to reach a detection. However, it is very unlikely that Proxima b is transiting \citep{Kipping2017}. A more promising avenue is to combine high-dispersion spectroscopy with high-contrast imaging \citep[HDS+HCI][]{Sparks2002, Riaud2007, Snellen2014, Snellen2015, Kawahara2014, luger2017}. \citet{Snellen2015} simulated observations with the E-ELT using optical HDS+HCI of a then still hypothetical Earth-like planet around Proxima, showing that detection of such planet would be possible within one night. Recently, \citet{Lovis2016} argued that if the new ESPRESSO high-dispersion spectrograph at the ESO Very Large Telescope (VLT) can be coupled with the high-contrast imager SPHERE, and the latter has a major upgrade in adaptive optics and coronagraphic capabilities, a detection of Proxima b is within reach. 

Since the next-generation of extremely large ground-based telescopes is at least 5 to 10 years away, the James Webb Space Telescope (JWST) could be a more immediate option to detect an atmospheric signature of Proxima b. Unfortunately, simple diffraction arguments tell us that the JWST is not large enough to spatially separate the planet from its host star - with a maximum elongation of  37 mas ($\sim$1$\lambda/D$ at 1 $\mu$m). Several studies \citep{Greene2016, deWit2016} show that atmospheric characterisation of super-Earths transiting small M-dwarf stars is within range of the JWST. However, the probability that Proxima b transits its host star is only 1.3\%.  Instead, \citet{Kreidberg2016} discuss the possibility of detecting the thermal phase curve with the JWST Mid-Infrared Instrument MIRI, using its slitless 5 - 12 $\mu$m Low Resolution (LRS) mode . Since the planet is expected to be tidally locked, 
the night-to-dayside temperature gradient will result in a variation in apparent thermal flux as function of orbital phase. Depending on the orbital inclination, and on whether the planet has an atmosphere (which affects the redistribution of absorbed stellar energy around the planet), variations of up to $\sim$35 ppm are expected in the LRS wavelength regime. They show that in the ideal case of photon-limited precision, one can indeed detect the phase variation over a planet orbit (11.2 days - 268 hrs). However, a particular concern
is the intrinsic variability of Proxima Cen - which is known to be a flare star. A month-long observation program of the MOST satellite \citep{Davenport2016} detected on average two strong optical flares a day.  Extrapolating their result to lower energies and mid-infrared wavelength implies that the star exhibits $\sim$50 flares a day at levels $>$500 ppm - an order of magnitude stronger than the expected phase variation amplitude of the planet.  Hence it will be challenging to discern the planet signal with such observations.

In this paper we discuss a new application to the cross-correlation technique used to probe exoplanet atmospheres \citep[e.g][]{Snellen2010}, targeting the 15 $\mu$m  CO$_2$ planet signal with the Medium Resolution Spectrograph (MRS) mode of MIRI. Importantly, it is unaffected by broadband flux variations of the host star. Furthermore, a detection of CO$_2$ would constitute conclusive evidence that Proxima b contains an atmosphere, and provide constraints to its temperature structure.  If Proxima b is indeed a terrestrial planet that formed at or near its current semi-major axis, it would have been subjected to the super-luminous phase of the star for up to 160 My \citep{Barnes2016}.  In this time it may have undergone ocean loss and a runaway greenhouse, or had all but the heaviest molecules in its atmosphere stripped early on, with the possibility of longer-term replenishment by volcanic outgassing over its 5 Gyr history \citep{Lammer2007, Ribas2016, Meadows2016}.   All of these evolutionary processes would have increased the likelihood that the planetary atmosphere currently contains CO$_2$.In Section 2 we describe the details of the method, of the MRS mode of MIRI, and present simulations, including atmospheric modelling. The results are presented and discussed in Section 3.

\section{A high-pass spectral filtering technique}

\begin{figure}
\label{example}
\plotone{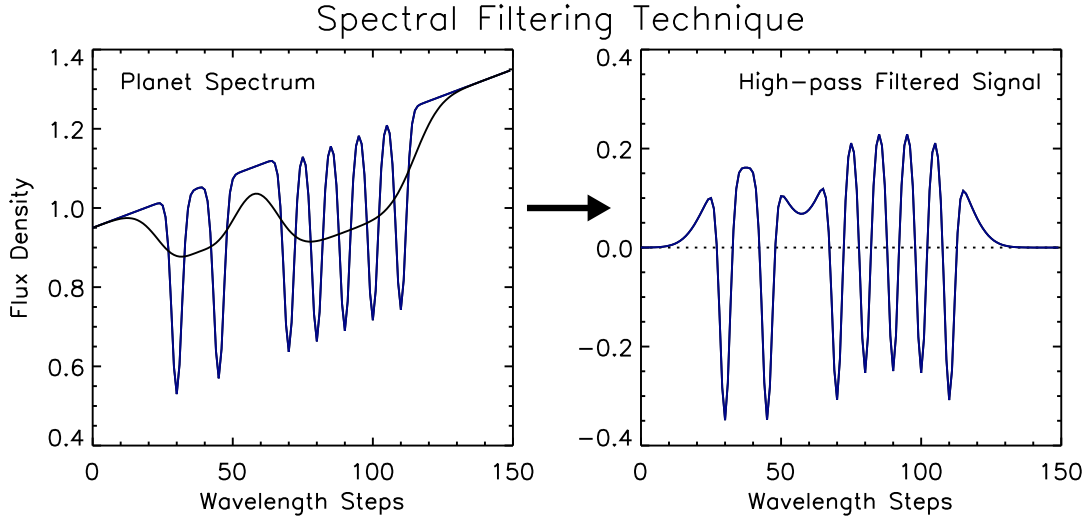}
\caption{Schematic representation of the high-pass spectral  technique. The left panel shows a toy-model planet spectrum (blue) with its low-resolution spectrum overlaid (black).  The right panel shows the high-pass spectrally filtered signal consisting of the difference between the high- and low resolution spectrum. Only the high-frequency components of the spectrum are preserved, making the high-pass filtered spectrum significantly easier to calibrate at a cost of information on the planet continuum flux level.}
\end{figure}

The key to any exoplanet atmospheric observation is the ability to separate planet signatures from the overwhelming flux of the host star.  The cross-correlation technique used to probe exoplanet atmospheres does not work in the case of Proxima b with the JWST, since the spectral resolution is not sufficient to use the change in the radial component of the planet orbital velocity to filter out the planet light \citep[e.g.][]{Snellen2010}, neither is the angular resolution sufficient to separate the planet from the star (in which case the HDS+HCI technique could be used). In this case we propose to use a new version of these techniques in which a spectral feature is targeted in the integrated planet+star spectrum that can be differentiated from the stellar spectrum and attributed to the planet. 
Generally, this would not work because it constitutes an absolute spectrophotometric measurement requiring instrumental calibration and knowledge of the stellar spectrum to a level significantly better than the planet signal. However, if the instrument spectral resolution is sufficiently high, the requirements on calibration and stellar spectrum knowledge can be significantly relaxed by probing the high-pass spectrally filtered signature instead of the absolute spectrophotometric signal. In this case the planet signal is the difference in flux density relative to a low-resolution mean. E.g. in the case of a molecular band composed of a series of distinct absorption lines, the high-pass filtered spectral signature consists of series of peaks and valleys in the spectrum (Fig. 1). This has the advantage that the planet signal is spread out over many pixels and consists of positive and negative high-frequency components.
In particular low-frequency components in the spectrum are notoriously difficult to calibrate due to required accuracies in the spectrum of the calibration source, and in stray-light and background corrections. However, these are not important in this case. 

We argue  that the spectral filtering technique will work well for the MRS mode of MIRI targeting CO$_2$ at 15 $\mu$m. In the case of Proxima b, its variations in the radial component of its orbital velocity of up to 50 km sec$^{-1}$ corresponds to a shift of $\pm$1 wavelength step, which may also be detected and subsequently constrain its orbital inclination. 

\subsection{MRS mode of MIRI}

The Medium Resolution Spectrograph (MRS) mode of the Mid-Infrared Instrument \citep[MIRI][]{Rieke2015, Wright2015} on JWST utilises an integral field spectrograph (IFS) which has four image slices producing dispersed images of the sky on two 1024x1024 infrared detector arrays, which provide R = 1300-3600 integral field spectroscopy over a $\lambda$ = 5 $-$ 28.3 $\mu$m wavelength range \citep{Wells2015, Labiano2016}.   The spectral window is divided in four channels covered by four integral field units: (1) 4.96-7.77 $\mu$m, (2) 7.71-11.90 $\mu$m, (3) 11.90-18.35 $\mu$m, and (4) 18.35-28.30 $\mu$m. 

Two grating and dichroic wheels select the wavelength coverage within these four channels simultaneously, dividing each channel into three spectral sub-bands indicated by $A$, $B$, and $C$ respectively. To obtain a complete spectrum over the whole MIRI band one has to combine exposures in the three spectral settings, A, B, and C. Since we are primarily interested in the 15 $\mu$m CO$_2$ feature, only one setting will be sufficient: $3B$ covering the 13.2 $-$ 15.8 $\mu$m range. Note that the same setting will deliver  the 1B (5.6 -- 6.7 $\mu$m), 2B (8.6 -- 10.2 $\mu$m) \& 4B (20.4 -- 24.7 $\mu$m) wavelength ranges for free. Of these, 2B is particularly interesting since it contains the ozone absorption feature. This is briefly discussed in Section 3.5. 

The IFS of Channel 3 consists of 16 slices (width = 0.39$''$), each containing 26 pixels (0.24$''$) providing a field of view of $\sim$ 6$'' \times 6''$. The spectrum is dispersed over 1024 pixels (1 pix = 2.53 nm = 52 km sec$^{-1}$) at a spectral resolving power of R$\sim$1790--2640 (168 - 113 km sec$^{-1}$). 

\subsection{Modelling Proxima b and its atmosphere}

\begin{figure}
\plotone{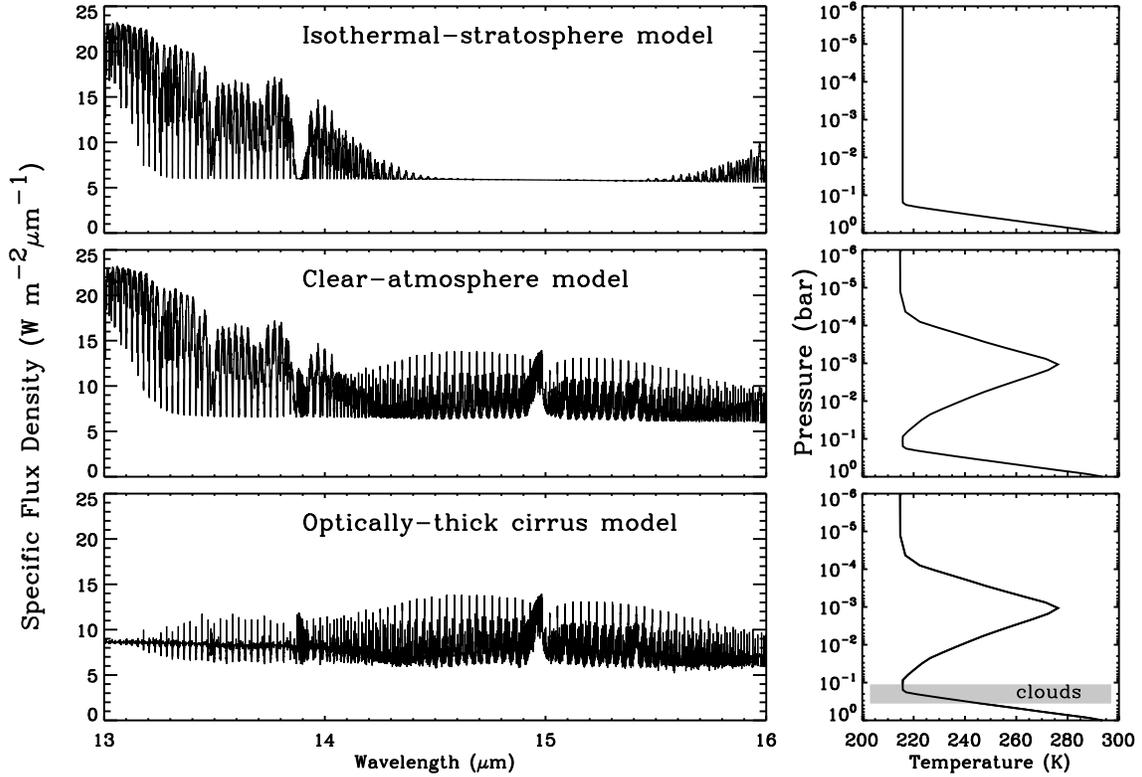}
\caption{Planet model spectra (see Sect. 2.2) assuming a standard Earth atmospheric model for temperatures and gas mixing ratios, with on the right the assumed T/p profile. The upper panel shows the case where the stratospheric temperatures were artificially made isothermal and equal to the tropopause temperature (isothermal stratosphere model), the middle panel shows a model for clear sky conditions (clear atmosphere model), and the lower panel shows a spectrum for a case with opaque high-altitude cirrus clouds (optically thick cirrus model).}
\end{figure}

Exoplanet Proxima b is found to orbit its host star in 11.186$^{+0.001}_{-0.002}$ days \citep{Anglada2016}. The amplitude of its radial velocity variations corresponds to a minimum mass of 1.27$^{+0.19}_{-0.17}$ M$_{\rm{Earth}}$. If the mean density of the planet is the same as that of the Earth, and its orbit is nearly edge-on, it will have a radius of $\sim$1.1 R$_{\rm{Earth}}$. Proxima Cen has an estimated mass of 0.123$\pm$0.006 M$_{\rm{Sun}}$, implying an orbital semi-major axis of 0.0485$^{+0.0041}_{-0.0051}$ AU, corresponding to a maximum angular separation of 37.5 mas. Proxima Cen has an effective temperature of $T_{\rm{Eff}}$= 3042$\pm117$ K, radius of 0.141$\pm$0.007 R$_{\rm{Sun}}$,  and bolometric luminosity of L=0.0017 L$_{\rm{Sun}}$ \citep{Doyle1990, Segransan2003, Demory2009}. 

Due to the close vicinity of the planet to its host star, it is generally assumed that Proxima b is tidally locked, meaning that the same dayside hemisphere is eternally facing the star. The planet effective dayside temperature will strongly depend on its Bond albedo and global circulation patterns. If due to atmospheric circulation the absorbed stellar energy is homogeneously distributed over the planet, and the Bond albedo is similar to that of Earth (A$_B$=0.306), the dayside equilibrium temperature of Proxima b is 235 K. If there is effectively no  circulation and the absorbed stellar energy is instantaneously reradiated, its observed dayside temperature could be as high as 300 K (and even up to 320 K for a moon-like albedo) . In the other extreme case in which the planet has an albedo such as Venus (A$_B$=0.9) with a very effective atmospheric circulation, its dayside effective temperature could be as low as 145 K. For the calculations below we assume a continuum brightness temperature of 280 K at 15 $\mu$m and a near-transiting orbital inclination, corresponding to a planet/star contrast ratio of 6$\times 10^{-5}$ at superior conjunction.

Simulated high-resolution emission spectra of Proxima b were generated by the Spectral Mapping Atmospheric Radiative Transfer (SMART) model assuming it to have an atmosphere such as Earth, using opacities from the Line-By-Line ABsorption Coefficient (LBLABC) tool \citep[both developed by D. Crisp; see][]{Meadows1996}.  The HITRAN 2012 line database \citep{Rothman2013} was used as input to LBLABC, which generates opacities at ultra-fine resolution (resolving each line with $>$10 resolution elements within the half-width) on a grid of pressures and temperatures that spans a range relevant to Earth's atmosphere.  Following this, SMART -- which has been extensively validated against moderate- to high-resolution observations of Earth \citep{Robinson2011} -- was used to simulate spectra at 5$\times$10$^{-3}$ cm$^{-1}$ resolution (corresponding to R $>$ 10$^5$ at the simulated wavelengths).

Our spectral simulations used a standard Earth atmospheric model for temperatures and gas mixing ratios \citep{McClatchey1972}, the spectra of which are shown in Figure 2.  To bound certain extremes in thermal emission, model runs were performed for both clear sky conditions (the `clear atmosphere model') and for an opaque high-altitude cirrus cloud \citep[located at 0.2 bar, near the tropopause][]{Muinonen1989}, called the `optically-thick cirrus model'.  Also, to explore a situation with large thermal contrast between the surface and stratosphere, a case where Earth's stratospheric temperatures were artificially made isothermal and equal to the tropopause temperature (210 K) was simulated (`isothermal stratosphere model').

\subsection{Simulated observations}

First, an estimate of the expected signal-to-noise (S/N) for Proxima Cen with MIRI is obtained from the beta version of the JWST exposure time calculator\footnote{http://jwst.etc.stsci.edu}. The 12 $\mu$m  and 22 $\mu$m flux densities of Proxima have been determined by the NASA Wide-field Infrared Survey Explorer (WISE) to be 924 mJy (m$_{\rm{W3}}$ = 3.838$\pm$0.015) and 278 mJy (m$_{\rm{W4}}$=3.688$\pm$0.025) respectively, which are fitted to a 3000 K Planck spectrum and subsequently interpolated to 13, 14, and 15 $\mu$m  flux densities of 816, 713, and 630 mJy respectively. These fluxes are fed to the exposure time calculator for
channel 3B of the MIRI MRS mode. A detector setup of 5 groups and fast readout gives an integration time of 16.6 seconds delivering an SNR of 200 for a total exposure time of 38.85 seconds.  This extrapolates to a S/N of  2000 per hour assuming that calibration uncertainties do not contribute to the noise.  

\begin{figure*}
\includegraphics{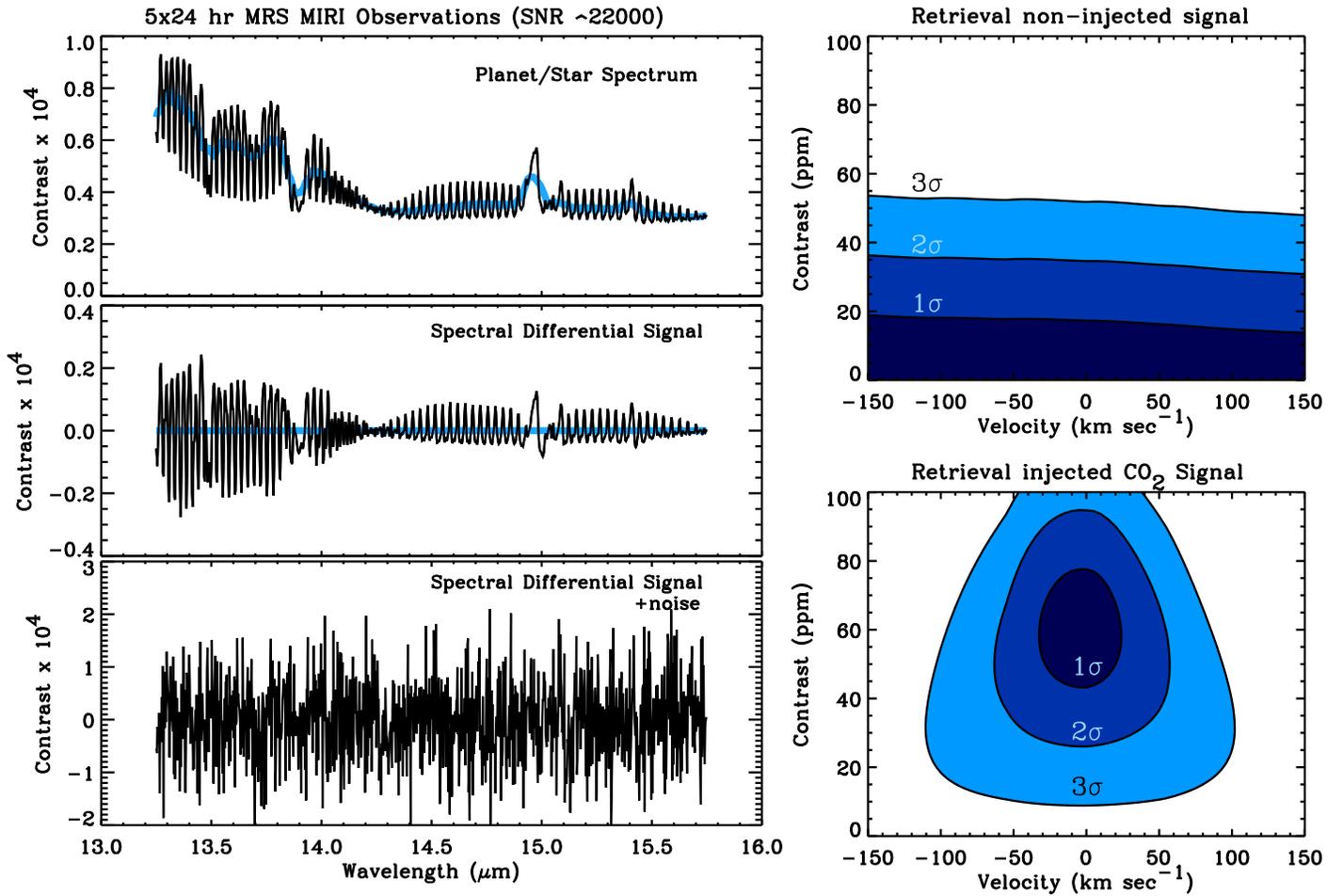}
\caption{\label{difspectrum} From model spectrum to simulated MRS MIRI observations assuming the clear-atmosphere model. The top-left panel show the model spectrum convolved to the resolution of the MRS 3B channel of MIRI, normalised relative to the average star flux. The middle-left panel shows its associated high-pass spectral signal and the lower-left panel with noise added as if Proxima was observed for 5$\times$24 hrs. The top-right panel shows the statistical 1, 2, and 3$\sigma$ confidence intervals when no CO$_2$ signal is present, pointing to a 3$\sigma$ upper limit of 50 ppm for the planet/star contrast. The bottom-right panel shows the same, but with the simulated CO$_2$ signal present - detected at $\sim$3.8 $\sigma$. 
The y-axis indicates the mean planet/star contrast of the (non-spectrally filtered) template spectrum between 13.2 -- 13.5 $\mu$m. Note that these simulations assume that both  the instrumental spectral response and the stellar spectrum at a wavelength-to-wavelength scale have been determined, which is likely to require extra deep observations at inferior conjunction that serve as a reference (see Section 3.2).}
\end{figure*}

The model planet spectra are smoothed to the spectral resolving power of R=2200  (the mean of the MRS 3B band) and subsequently binned to the wavelength steps of the 3B MRS channel. The resulting clear-atmosphere model spectrum  normalised by the stellar spectrum (which for clarity is assumed to be featureless) is shown in the top panel of Fig. \ref{difspectrum}. Since the spectral filtering technique is only sensitive to the high-frequency signals, the low-frequency components are removed by subtracting a 25 wavelength-step sliding average (a rather arbitrary width) from the spectrum, resulting in the spectral differential spectrum shown in the middle panel. Subsequently, random noise is added to this differential spectrum at a level expected for the total simulated integration time, as shown in the bottom panel of Fig. \ref{difspectrum}.

At this stage it is determined at what statistical significance level the differential model spectrum is preferred to be present in the data  with respect to pure noise. This is done by calculating the chi-squared of the observed spectrum, with its sliding average and differential model spectrum removed - for a range of planet/star contrasts and radial velocities.  The minimum chi-squared is assigned as the best fit and the $\Delta \chi^2$ interval is used to determine the statistical uncertainties of a possible CO$_2$ detection. These simulations were repeated for the three different models, and for a range star/planet contrasts corresponding to different orbital phases or different effective dayside temperatures. 
 
\section{Results and Discussion}

\subsection{Detectability}

Our simulations show that the 15 $\mu$m CO$_2$ high-pass filtered signal of the Earth-mass planet can be detected within a limited amount of observing time. The MRS mode of MIRI at the JWST will in 24 hours integration time (excluding overheads) deliver a R=1790--2640 spectrum of Proxima Cen between 13.2 and 15.8 $\mu$m at a S/N of $\sim$10,000 per wavelength step (assuming photon noise). This corresponds to a 1$\sigma$ contrast limit of $\sim 1 \times 10^{-4}$. While the high-frequency features in the filtered planet spectrum are typically at a 1--3$\times 10^{-5}$, there are about 100 within the targeted wavelength range - combining to a detection at a $\sim$2$\sigma$ level. It means that while the continuum planet/star contrast is at a level of 6$\times 10^{-5}$ ($\sim$0.6$\sigma$ per wavelength step),  the combined spectrally filtered signal over the 3B band is about a factor 3--4 higher. 
The top-right panel of Fig. \ref{difspectrum} shows the statistical confidence intervals for 5$\times$24 hrs of observations if no CO$_2$ signal is present, while the bottom-right panel shows the same, but then with the clear-atmosphere CO$_2$ model spectrum for a face-on planet injected, indicating it can be detected at nearly 4$\sigma$ within this exposure time. Hence, while the individual CO$_2$ features are not visible in the simulated spectrum, their combined signal can be clearly detected. Results are very similar for the Isothermal Stratosphere model and the Optically-Thick Cirrus model. 

\subsection{Important prerequisites}

\subsubsection{The stellar spectrum and its variability}

We have made several assumptions that are vital for the high-pass spectral filtering technique to succeed in detecting CO$_2$ in Proxima b.  First, it is assumed that the high-frequency components of the spectrum of the host star itself are perfectly known. Although low-resolution (R=600) mid-infrared spectra of M-dwarfs taken with the Spitzer Space Telescope \citep{Mainzer2007} seem featureless, Phoenix model spectra\footnote{https://phoenix.ens-lyon.fr/Grids/BT-Settl/CIFIST2011\_2015/SPECTRA/} \citep{Allard2012} show that the 13.2-15.8 $\mu$m wavelength region of an M5V dwarf star harbours thousands of H$_2$O lines, collectively resulting in wavelength-to-wavelength variations of $\sim$ 2\% in the MIRI MRS spectrum.  It means that these features need to be calibrated to better than a relative precision of 1\% for them not to interfere with the planet CO$_2$ signal and not to act as an extra noise source. The star is also expected to have numerous but much weaker lines from hot CO$_2$  in this MRS band, resulting in wavelength-to-wavelength fluctuations of a few times 10$^{-4}$ - hence which need to be calibrated to a relative precision of $\sim$10\%.  It is rather unlikely that spectral modelling could be sufficient to calibrate the high-frequency components of the mid-infrared spectrum to $\le 10^{-4}$. This means that  a deep stellar spectrum  may need to be obtained near or at inferior conjunction when the contribution from planet emission from the system is the smallest, which would limit the sensitivity of this method for low orbital inclinations and would take as much time as the observations at superior conjunction. 

Since the data at superior and inferior conjunction must be taken at least $\sim$5 days apart, variability of spectral features should also be below $\le 10^{-4}$ on this time scale. We used archival data of Proxima from the UVES spectrograph (R=100,000) at the Very Large Telescope separated by 4 days (October 10 \& 14, 2009) to assess the optical variability of the star. For each of the two nights, a few dozen spectra were combined and the 868 -- 878 nm wavelength range extracted which is dominated by hundreds of TiO lines but is free of telluric lines. The averaged spectrum was subsequently convolved with a Gaussian to mimic the resolution of MIRI and subsequently binned to match its wavelength steps (in $\Delta \lambda / \lambda$). After dividing out a linear trend with wavelength, the standard deviation of the ratio of the resulting spectra of the two nights is 4$\times 10^{-4}$. Since these data are possibly limited by flat fielding uncertainties, and variability in the mid-infrared is expected to be lower, this result is encouraging.

\subsubsection{Instrument calibration}

Another prerequisite is that the spectral responses of the MRS pixels of MIRI can be adequately calibrated. Neither absolute flux calibration nor the low-frequency spectral response are important, but the sensitivity of one wavelength relative to the next is crucial - e.g. the  spectral pixel-to-pixel calibration of the flat field. Potentially challenging is fringing, a common characteristic of infrared spectrometers. It is caused by interference at plane-parallel surfaces in the light-path of the instrument.  Experiences with data from ISO and Spitzer show that it can be removed down to the noise level \citep[e.g.][]{Lahuis2000}. \citet{Wells2015} have characterised the fringing of the MIRI detectors in the laboratory and identify three fringe components with scale lengths (in wave number) of 2.8, 0.37 and 10-100 cm$^{-1}$, originating from the detector substrate, dichroic, and fringe beating respectively. The planet CO$_2$ features also show a regular pattern, but with a characteristic scale length of $\sim$1.6 cm$^{-1}$, which fortunately is significantly different from these fringe components. A potentially unwelcome source of error may be fringing in combination with dithering. Small residuals left over after defringing, combined from different dither positions, may be challenging to calibrate. 

In the signal-to-noise calculations presented above we assumed that the instrument calibration is perfect. For it not to add an extra source of noise, the wavelength-to-wavelength precision of the flat fielding and fringe removal must be $\le 10^{-4}$. If this level can be reached for individual IFS pixel, a tailored dithering strategy will subsequently push the calibration noise to below 10\% of the noise budget for a 24h observation. A single observation will have the star light mostly distributed over 2 slices $\times$ 2 pixels in the IFS, and by moving the star stepwise over the field of view of the IFS about 80 independent positions can be obtained. This will reduce the calibration noise by a factor $\sqrt{2\times2\times80} \approx 15$. 

\subsubsection{Planet spectrum}

In addition, temporal stellar atmospheric disturbances can modify the chemical composition of a planet atmosphere, meaning that flares could temporarily change the CO$_2$ abundance of Proxima b. Such effects has been investigated by \citet{Venot2016}, who find that although the abundances of some chemical species can be significantly altered deep in to the atmosphere ($\sim$1 bar),   CO$_2$ is expected to only be affected very high in the atmosphere at $< 6 \times 10^{-4}$ mbar - suggesting that the 15 $\mu$m CO$_2$ will hardly be affected. 

In principle, the observations could also be sensitive to other planets in the system. Since such planet would likely be in a significantly wider orbit and be colder, the expected signal would be smaller. The CO$_2$ signal of such hypothetical planet could be distinguished from that of Proxima b since its superior conjunctions would occur on different epochs.

\subsection{Phase variations}

\begin{figure}
\vspace{-3cm}
\includegraphics[angle=-90,scale=0.65]{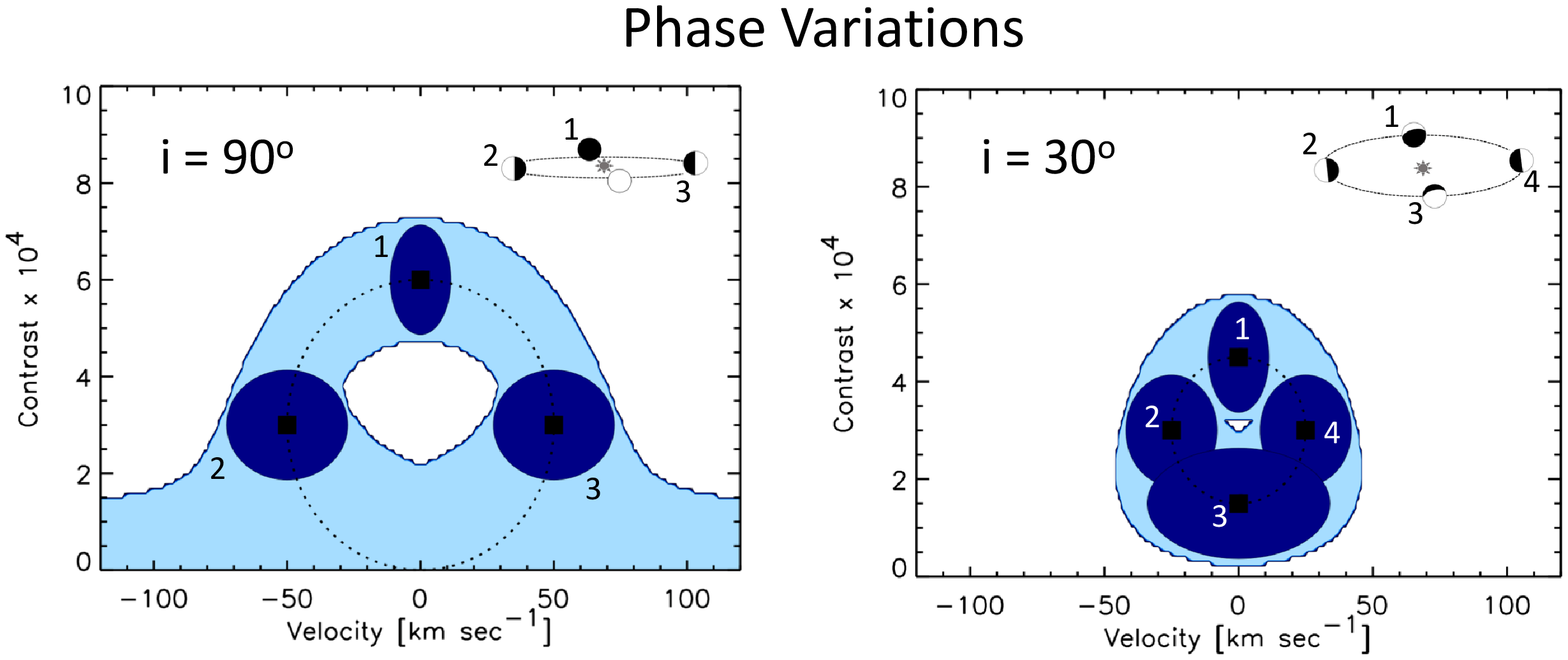}
\vspace{-3cm}
\caption{Examples of phase variation observations of the 15 $\mu$m CO$_2$ feature of Proxima b. The left shows the expected variation for an edge-on orbit, with the light blue regions indicating the 1$\sigma$ confidence interval for a 9$\times$24 hr observation at a given orbital phase (hence nearly 650 hrs of observations in total - significantly more than the simulations presented in Fig. 3). The three dark-blue regions indicate the 1$\sigma$ confidence intervals for particular observations at an orbital phase of $\phi$=0.5 (superior conjunction), 0.25 and 0.75 - showing the variation in radial velocity and contrast.  The right panel shows the same but for an inclination of $i = 30^o$, resulting in a smaller variation in contrast and radial velocity as function of phase. In both cases it is assumed that all thermal flux originated from the dayside hemisphere of the planet with an effective temperature of 280 K.\label{combined_signal}}
\end{figure}

A  detection of CO$_2$ will provide us with both the strength of the planet signal and its radial velocity. These can be used to constrain the orbital inclination of Proxima b. An example of such observation is shown in Fig. \ref{combined_signal}, showing in the left panel the expected variation in contrast and radial velocity and their uncertainties for 9$\times$24 hrs exposures for each three measurements at orbital phase $\phi$=0.25, 0.5, and 0.75 - assuming an orbital inclination of near 90$^o$. These represent significantly longer exposures than that presented in Fig. 3.  The right panel shows the same but for an inclination of i = 30$^o$, resulting in a smaller variation in contrast and radial velocity as function of phase. The observations at inferior conjunction may need to serve as a reference for the stellar spectrum (see Sect. 3.2.1). In both cases it is assumed that all thermal flux originated from the dayside hemisphere of the planet with an effective temperature of 280 K. 

Since the strength of the signal at $\phi$=0.25 and 0.75 can be up to a factor two lower than that at $\phi=0.5$, calibration of the instrumental response and the stellar spectrum will  be even more important. If the orbital inclination is low, the planet will never be seen entirely face-on, reducing the maximum signal at superior conjunction. However, it would also mean that the mass of the planet is higher, meaning that the planet radius could be larger than assumed above (in particular if such more massive Proxima b is volatile rich), possibly counteracting the reduction in expected planet surface brightness.  

\subsection{Atmospheric characterisation}

We performed our MRS MIRI simulations for three different atmospheric model spectra, 1) for a planet with an isothermal stratosphere, 2) for a planet with an inversion layer and clear atmosphere, and 3) with inversion layer combined with optically thick cirrus. Independent of which model we use, the increase in S/N over that expected for one wavelength step are very similar for all models at a factor 3--4. We also experimented by using the Earth transmission spectrum instead, which is similar to the isothermal stratosphere spectrum, but with significant differential signal at the heart of the CO$_2$ band. This provides a S/N increase of a factor of $\sim$5 over the S/N at at single wavelength step, which can be treated as an upper limit to the differential gain. 

Retrieving an injected signal from one model using one of the other spectra results in a significant decrease in signal to noise - not surprising since a large number of the spectral features appear as either emission or absorption in the models. The brightness temperature of the planet atmosphere at a certain wavelength roughly corresponds to the atmospheric temperature at the $\tau = 1$ surface. In the center of the strongest CO$_2$ lines, where the opacity is greatest, we probe the atmosphere at the highest altitudes. Therefore, in the case of a strong thermal inversion (the clear atmosphere and the optically-thick cirrus models) the atmosphere will be warmer at such low pressure - resulting in emission lines in stead of absorption lines when the atmosphere is cooler at higher altitudes.   This implies that a detection will also constrain the temperature structure of the upper atmosphere, giving additional insights in high-altitude atmospheric processes. It will not just merely be a detection of the planet atmosphere, which can be compared with theoretical models. E.g. \citet{Segura2005} argue that Earth-like planets orbiting M-dwarfs are likely to have relatively cool, bordering on isothermal stratospheres -- even with O$_3$ present.

Several features from other molecules are present in the 13.2 - 15.8 $\mu$m wavelength range, such as C$_2$H$_2$ at 13.7 $\mu$m and HCN at 14 $\mu$m, which may be included in the atmospheric spectral template if needed (and if expected to be present in the planet atmosphere). 

\subsection{Prospects of detecting ozone}

When  CO$_2$ is targeted in the MRS 3B band,  the 1B, 2B, and 4B bands are observed simultaneously. Interestingly, the 2B band, ranging from 8.6 to 10.3 $\mu$m covers the 9.6$\mu$m ozone band. While CO$_2$ in the atmosphere of Proxima b may be likely, if the planet did undergo ocean loss early it its history to generate a massive O$_2$ atmosphere \citep{Luger2015}, it is also possible that O$_3$, photochemically-produced from the O$_2$, is also present in higher abundances than seen on Earth \citep{Meadows2016}.  O$_3$ is of course also of high interest as it can be used as a proxy for the O$_2$ biosignature from a photosynthetic biosphere.  Detection of O$_3$ with JWST would therefore provide an intriguing first hint that life might be present on an extrasolar planet, although O$_3$ production via abiotic O$_2$ from ocean loss would first have to be ruled out. 

 The expected S/N per wavelength step in a 24 hour observation is 2$\times 10^4$, about a factor of two higher than in the 3B band because of the high stellar flux. On the other hand, the expected continuum planet/star contrast is a factor $\sim2$ lower compared to that expected at 15 $\mu$m. Unfortunately, the individual lines within the 9.6 $\mu$m band are more tightly packed than the lines in the 15 $\mu$m CO$_2$ band, i.e. the ozone band is not fully resolved at the MRS resolution of R=2800 in this wavelength range.
This means that if ozone is present in the atmosphere of Proxima b, it's spectral differential signal will be about a factor 3--4 smaller than that of CO$_2$.  We estimate that in the best case one could expect a 2$\sigma$ result in 20 days of JWST observing.  We also note that the prerequisite for spectral calibration is also more stringent by a factor 2 compared to the CO$_2$ case. Hence, although observations of ozone come for free when the CO$_2$ band is targeted, it is unlikely this could result in a firm detection and is probably beyond the limit of what the JWST can achieve. 

\subsection{Strategy for proof of concept and other prospects}

We envisage two ways to show proof of concept for the high-pass spectral filtering technique with the MRS mode of MIRI. First, the method can be used on exoplanet targets with significantly higher planet/star contrasts. For example, a T=1000 K hot Jupiter orbiting a solar type star will have a 15 $\mu$m contrast of 10$^{-3}$, a factor 20 higher than Proxima b. It means that for a host star whose 15 $\mu$m flux is 40 times (4 magnitudes) fainter than Proxima b, CO$_2$ will still be detected 10x faster. Also, one could aim for cool Neptunes or super-Earths orbiting nearby M-dwarfs, such as Gliese 687b. If the spectrum of this particular planet exhibits a CO$_2$ absorption feature it can be detected a factor 4 faster than in the case of Proxima b.  From a theoretical point of view it will be important to identify those planets that are expected to have CO$_2$ in their atmospheres and select those with the most favourable stellar magnitudes and planet/star contrasts. 

Ultimately, one should target Proxima itself, gradually increasing the integration time and validating at each step that the expected S/N limits are being reached. Detecting CO$_2$ in the atmosphere of Proxima b will be a major step forward in our quest for potential habitats and signs of extraterrestrial life. 

The detection of specific spectral features expected in a planet atmosphere becomes orders of magnitude more powerful if the planet can also be angularly separated from the planet \citep[e.g.][]{Snellen2015}. In such case the stellar spectrum can be effectively removed from all pixels in the IFU (since it is identical everywhere), after which the residual spectra can be searched for the planet features. Hoeijmakers et al. (2017; in prep.) show that this technique is very effective for the SINFONI and OSIRIS IFU spectrographs, located at  the VLT and Keck Telescopes respectively, which have spectral resolving powers similar to that of MIRI (and the NIRSPEC IFU). If we could point the JWST directly at $\alpha$ Centauri A using the MIRI MRS, only 1--1.5$''$ away the starlight is at the level of that of Proxima implying that an Earth-size planet in the habitable zone of $\alpha$ Cen A could possibly be detected in 24 hrs. Unfortunately, $\alpha$ Centauri A will saturate the MIRI detectors within a small fraction of a second, irreversibly damaging the instrument.  Possible ways to mitigate this issue need to be investigated.

\section*{Acknowledgments}

Snellen acknowledges funding from the European Research Council (ERC) under the European Union's Horizon 2020 research and innovation programme under grant agreement No 694513, and from research program VICI 639.043.107, which is financed by The Netherlands Organisation for Scientific Research (NWO).  D\'esert acknowledge funding from the European Research Council (ERC) under the European Union's Horizon 2020 research and innovation programme (grant agreement nr 679633; Exo-Atmos).
Robinson and Birkby gratefully acknowledge support from the National Aeronautics and Space Administration (NASA) through the Sagan Fellowship Program executed by the NASA Exoplanet Science Institute. Support for this work was provided in part by NASA through Hubble Fellowship grant HST-HF2-51336 awarded by the Space Telescope Science Institute, which is operated by the Association of Universities for Research in Astronomy, Inc., for NASA, under contract NAS5-26555. Meadows and Robinson are members of the NASA Astrobiology Institute's Virtual Planetary Laboratory Lead Team, supported by NASA under Cooperative Agreement No. NNA13AA93A.
\bibliography{bibtexIgnas}


\end{document}